\documentclass[24p]{article}
\usepackage[dvips]{graphicx}

\begin{document}
\baselineskip=18pt
\pagestyle{plain}
\setcounter{equation}{0}

\begin{center}
\begin{LARGE}
Derivation of Dielectric Model of Confimenent in QCD
\end{LARGE}
%}% Force line breaks with \\

\begin{large}
\vspace{0.8cm}
%\author{
R. Fukuda 

\vspace{0.8cm}
%\protect\\
Department of physics,
Faculty of Science and Technology,
 Keio University, \protect
\\
Hiyoshi 3-14-1, Yokohama 223-8522, Japan
\end{large}

\vspace{1.5cm}
\begin{large}
{\bf Abstract}
\end{large}
\end{center}

\vspace{0.3cm}
\noindent
After the gauge invariant gluon condensation, gluons
remain as massless excitations. When an effective theory 
desribing the condensation and the excitaion is constructed, 
 a constraint has to be imposed 
 for the vacuum to be stable.
 The constraint implies the perfect dia-electricity and 
 assures the solution of color flux tube when qurks are introduced. 
Thus the dia-electric model of Kogut-Susskind \cite{Susskind}
 and 't Hooft \cite{tHooft} 
are derived by the stability of the condensed vacuum.

\vspace{0.4cm}
\noindent
PACS nimber;~~~11.15-q, 12.38Aw, 12.38Lg.

\vspace{0.4cm}
\noindent
\begin{center}
{\bf Introduction};~
\end{center}

Understanding of the confinement mechanism 
based on QCD Lagrangian remains as a fundamental problem in the 
strong interaction regime. Besides lattice QCD \cite{Lattice}, 
active researches have been performed 
along the line of the dual superconductivity 
 with the 
abelian gauge fixing \cite{abel,Ripka}, infrared structure
 of the gluon propagators \cite{Gribov},  
 e.t.c.. In continuum QCD, the key point will be how to 
extract the effect of the gluon condensation.
\\~~
In this letter, we propose an approach to construct the
 effective Lagrangian and   
derive the dielectric model of Kogut-Susskind \cite{Susskind}
 and 't Hooft \cite{tHooft}. We asssume that 
  gluons condense in gauge invariant form, so 
the gauge invariance is not broken in the vacuum and 
gluons which are excitations above the condensed vacuum remain massless.  
 Because of masslessness,
 they may condense further in color singlet form. Therefore, when we 
construct the effective 
Lagrangian describing the interaction between 
the condensation and excitation field, 
 non-trivial conditions have to be imposed to assure the 
 the stability of the vaccum. 
Remarkably,
 these conditions imply the perfect dia-electricty of the vacuum 
and guarantee the 
existence of the color flux tube of infinite length when quarks are introduced. 
 As for the gluon condensation, the magneteic type condensation
 $<\!\!G_{\mu\nu}^2 \!\!>>0$
has been established \cite{Schifman}
 but the advantage of the present work  
is that we do not have to ask the precise mechanism of the condensation, 
the magnetic monopole\cite{monopole} or
 the gluon pairs\cite{Fukudakugo}
 e.t.c.. 

\begin{center}
{\bf Effective Lagrangian ${\mathcal{L}}_{\rm eff}$};~
\end{center}

To measure the singlet condensation, any gauge invariant operator can be used but 
we choose for simplicity the normal oredred form of $G_{\mu\nu}^2$; 
$\phi(x)={\mathcal{N}}G_{\mu\nu}^{2} (x)
=G_{\mu\nu}^{2} (x)-<\!\!0|G_{\mu\nu}^{2} (x)|0\!\!>$.   
($|0\!\!>$ is the normal vacuum and 
 the dimensional regularization is adopted.)
%BBB which is gauge invariant 
%BBB and gives zero for the subtracted term in four dimension.) 
 With $|0\!\!>_c$ the condensed vacuum, we know that 
 $\mbox{}_{c}\!\!<\!\!0|{\mathcal{N}}G_{\mu\nu}^2 
|0\!\!>_c \equiv \phi_c >0$.
 We regard gluon fields 
$A^a_{\mu}(x)$ and $\phi(x)$ as independent fields. 
(The general formalism to achieve this 
 is shown below.)
As in the case of the free energy of the 
phenomenological theory of Ginzburg-Landau, the low energy 
effective Lagrangian ${\mathcal{L}}_{\rm eff}$ is obtained by 
the hydrodynamic expansion.
 It agrees with the expansion according to the operator dimension 
 in $A^a_{\mu}(x)$. By gauge invariance  
and keeping lowest non-trivial terms, we get
%BBB one gets 
\begin{eqnarray}
&&
{\mathcal{L}}_{\rm eff}=
\partial_{\mu}\phi(x)\partial^{\mu}\phi(x)/2-V(\phi(x))
\nonumber
\\&&~~~~~~~~~~~~
-\epsilon(\phi(x))G^{2}_{\mu\nu}(x)/4
\equiv {\mathcal{L}}_{\phi}+{\mathcal{L}}_{\epsilon,A},
\label{GL}
\\&&
G^{a}_{\mu\nu}(x)=\partial_{\mu}A^a_{\nu}(x)-\partial_{\nu}A^a_{\mu}(x)
+gf^{abc}A^b_{\mu}(x)A^c_{\nu}(x).
\nonumber
\end{eqnarray}
Here, $g$, $f^{abc}$ is the coupling constant and the structure constant of QCD, 
respectively. 
The effect of the condensation on the excitation appears as a dielectric 
factor $\epsilon(\phi)$. 
Assuming $\phi_c >0$, define $V(0)=0$, $V(\phi_c )<0$.  We require $\phi=0$ $(\phi=\phi_c )$ 
is the maximum (minimum) of $V(\phi)$; 
 $V'(0)=V'(\phi_c )=0$, $V''(0)<0$,~$V''(\phi_c )>0$.  
When $\phi=0$, 
${\mathcal{L}}_{QCD}$
 has to be recovered, so $\epsilon(0)=1$. 
 For all $\phi$, $\epsilon(\phi)$ has to be 
positive in order for ${\mathcal{L}}_{\epsilon,A}$ to be a sensible theory. 
The gauge fixing term and resulting ghost fields are not 
written excplicitly.  
Our conclusions will not change if these are introduced 
since only a gauge invariant degree $\phi(x)$ is added to the ordinary QCD, 
where we know how to extract physical sectors \cite{KugoOjima}.  

\begin{center}
{\bf Stability of the condensed vacuum};~~
\end{center}

Below, the fluctuation of the condensation is neglected, so $\phi(x)$ is 
treated as a c-number field. 
 In order for ${\mathcal{L}}_{\rm eff}$ to describe a consistent theory,  
i.e. $\phi=\phi_c$ actually realizes the lowest energy state of
 ${\mathcal{L}}_{\rm eff}$, 
%BBB is actually realized by , 
we have to require first of all that excitation fields
 $A^a_{\mu}(x)$ do not condense any more when $\phi=\phi_c$. 
If $\epsilon(\phi_c )\neq 0$, 
${\mathcal{L}}_{\epsilon,A}$ is proportional to ${\mathcal{L}}_{QCD}$, 
 so $A^a_{\mu}(x)$ condenses in the color singlet channel. In that case 
$\phi$ has to be redefined, and after that $\epsilon(\phi_c )=0$ is  
satisfied. Such a perfect dia-electricity of the stable vacuum
 is specific to QCD; the gauge invariance is preserved after the condensation. 
 Note that if the excitation 
acquires a mass gap, as in the case of the superconductor, such an instability 
never occurs.  
Assuming the homogeneous case $\phi=$ constant, we require that the energy 
is indeed increased for $\phi\neq\phi_c$. 
 Consider
 the trace of the energy momentum tensor in n-dimension by using ${\mathcal{L}}_{\rm eff}$;
$\Theta_{\mu}^{\mu}=4V(\phi)+\epsilon(\phi)(n-4)G_{\mu\nu}^2$. 
 In calculating the anomalous term for  
 $n\rightarrow 4$, we have to define ${\mathcal{N}}G^2_{\mu\nu}$, 
 by subtracting the expectation value
 taken by the perturbative vacuum of the ordinary QCD. 
In this conection, notice that our origin of the energy
 is that of the perturbative vacuum of QCD,
 for example we have defined $V(\phi=0)=0$. 
Thus the situation is that the whole system evolves
 with ${\mathcal{L}}_{\rm eff}$
 while the normal order has to be defined by the perturbative vacuum of 
${\mathcal{L}}_{QCD}$. In such a case, it is convenient to 
adopt the interaction picture with 
${\mathcal{L}}_{\epsilon=1,A}={\mathcal{L}}_{QCD}$ as the 
``free Lagrangian", and the rest as the ``interaction part",
 the latter being converted into the state vector.
 We can minimize the expectation value 
of $\Theta_{\mu}^{\mu}$ in the interaction picture thus defined.
(See below for details. )
Now the operator evolves by ${\mathcal{L}}_{QCD}$,
 so the trace anomaly of QCD holds;  
\begin{equation}
\Theta_{\mu}^{\mu}(x)=
%-\partial_{\mu}\phi(x)\partial^{\mu}\phi(x)+
4V(\phi)
+\epsilon(\phi)(\beta(g)/2g){\mathcal{N}} G^{a~\!2}_{\mu\nu}(x),
\label{anomaly}
\end{equation}
Here $\beta(g)$ is the usual $\beta$ function
 with $g$ the renormalized coupling constant. 
 We know that the excitation field 
 $A^a_{\mu}$ condenses with the amount 
\begin{equation}
 B=(\beta(g)/8g)
\mbox{}_{c}\!\!<\!\!0|{\mathcal{N}}G^{a~\!2}_{\mu\nu}(x)|0\!\!>_c ~\!\!<0. 
\label{B}
\end{equation}
With fixed $\phi$, the energy density (measured from the normal vacuum)
 of such a state is given by $e=V(\phi)
+\epsilon(\phi)B$. Here we have used the
 relation valid for any homogeneous vacuum;~ 
$<\!\!\Theta_{\mu\nu}(x)\!\!>=g_{\mu\nu}e$, so $<\Theta_{\mu}^{\mu}>=4e$.  
We require the above condensation does not lower the energy of the 
state $\phi=\phi_c$, otherwise $|0\!\!>_c$ is unstable. 
Here we note that 
$B$ introduced in (\ref{B}) is the energy 
 of condensed vacuum $|0>_c$, since it is 
calculated by ${\mathcal{L}}_{QCD}$, therefore 
 it is nothing but $V(\phi_c )$;   
%BBB both $V(\phi_c )$ and $B$ are the ground state
%BBB energy measured by $\phi$ and the 
%BBB codensation of the excitation field $A^a_{\mu}$ respectively.  
%BBB ${\mathcal{N}}G_{\mu\nu}^2$.
%BBB Thus  
 $V(\phi_c )=B$. In this way,  
the stability of $|0>_c$ requires 
\begin{equation}
B<V(\phi)+\epsilon(\phi)B,
\label{general}
\end{equation}
A remarkable fact is that the stability condition (\ref{general}) 
 assures the existence of the solution of color flux tube of infinite length 
 when quarks are introduced.
Near $\phi=\phi_c$, 
let us put $\phi=\phi_c +\Delta \phi$ in (\ref{general})  
and expanding $V(\phi)$ and $\epsilon(\phi)$ 
up to $(\Delta \phi)^2$. One arrives at  
\begin{equation}
\epsilon(\phi_c )=\epsilon'(\phi_c )=0,~~~~~~~V''(\phi_c )+
\epsilon''(\phi_c )B>0
\label{Ve}
\end{equation}
In case $\epsilon(\phi)$ is parametrized as 
$\epsilon(\phi)=C(\phi-\phi_c )^{2\alpha}$ with $C>0$ near $\phi=\phi_c$, then 
$\alpha\geq 1$.
Expanding in $\phi$ for small $\phi$, one obtains from (\ref{general})   
\begin{equation}
\epsilon'(0)=0,~~~~~V''(0)+\epsilon''(0)B>0.
\label{00}
\end{equation}
By $V''(0)<0$ and $B<0$, we get $\epsilon''(0)<0$.    
 Thus near $\phi=0$, 
$\epsilon(\phi)$ behaves as $1+a\phi^2 $ with $a<0$.

The inequality (\ref{general}) becomes an  
equality for $\phi=\phi_c$ and for $\phi=0$.
 This should be the case 
since both represent $|0\!\!>_c$; for $\phi=0$ the excitation $A^a_{\mu}$ 
condenses producing $B$.  
%BBB right-hand side of (\ref{general}) represents the same vacumm
%BBB state $|0\!\!>_c$ of QCD. 
 However we have to select $\phi=\phi_c$ as $|0\!\!>_c$, because our starting
 definition of $A^a_{\mu}$ 
 is that it represents the exitation field without the condesned part. 
If $\phi=0$ is selected, the role
 of $\phi$ and ${\mathcal{N}}G_{\mu\nu}^2$ is interchanged, 
 so we have to restart our discussions.
 We conclude that 
$\phi=\phi_c$ is the lowest state of energy for $0<\phi<\phi_c$,
 without ${\mathcal{N}}G_{\mu\nu}^2$ condensing anymore, since the condensation  
 does not lower the energy. 
 When $\phi=\phi_c$, we have no observable effect of 
${\mathcal{N}}G_{\mu\nu}^2$ due to $\epsilon=0$. (See below for more details.) 
Thus our effective Lagrangian is (\ref{GL}), with $V(\phi)$ and $\epsilon(\phi)$ 
satisfying (\ref{general}).  
Before showing the tube-like solution, 
we discuss two subjects whose results have been utilized above. 

\begin{center}
{\bf  Constructing ${\mathcal{L}}_{\rm eff}$};~
\end{center}

We present first the method of obtaining ${\mathcal{L}}_{\rm eff}$
by the Fourier tranform and its inverse, 
 with $\phi$ and $A^a_{\mu}$ being treated as independent fields. 
Let $\Psi[A^a_{\mu}]$ 
 be an arbitrary functional of $A^a_{\mu}(x)$ and start from 
\begin{equation}
Z=\int [dA^a_{\mu}]
\Psi[A^a_{\mu}]\exp{\rm i}\int
{\mathcal{L}}_{QCD}
d^4 x
\label{QCD}
\end{equation}
Let $O(A^a_{\mu}(x))$ be a gauge invariant 
operator and insert a functional identity 
$\int[d\phi]\int[dJ]\exp{\rm i}\int J(x)(O(A^a_{\mu}(x))-\phi(x))d^4 x$ 
and write $\Psi[A^a_{\mu}]$ by a functional Fourier transform $\Psi[j^a_{\mu}]$.  
%BBB by $\int[dj^a_{\mu}]\Psi[j^a_{\mu}]\exp{\rm i}\int
%BBB j^a_{\mu}(x)A^{a\mu}(x)d^4 x$. 
Then the integration over $A^a_{\mu}$ is done 
with ${\mathcal{L}}_{QCD}+j^a_{\nu}A^{a\mu}+JO(A^a_{\mu})$ in the exponential. 
Writing the result as $\exp{\rm i}W[j^a,J]$, 
\begin{eqnarray}
Z &=&
 \int[dJ]\int[dj_{\mu}^a ]
 \int[d\phi]\Psi[j^a_{\mu}]
 \exp{\rm i}W[J,\phi,j]
\nonumber\\&\equiv&
\int[dA^a_{\mu}]\int[d\phi]\Psi[A^a_{\mu}]
\exp{\rm i}\Gamma[\phi,A^a_{\mu}].
\label{GammaA}
\end{eqnarray}
is obtained. Here $W[J,\phi,j]=W[J,j_{\mu}^a ]-\int J(x)\phi(x)d^4 x$. 
In arriving at 
(\ref{GammaA}), $\Psi[j^a_{\mu}]$ has been transformed back to 
 $\Psi[A^a_{\mu}]$, and we have done integrations over $j^a_{\mu}$ and $J$. 
Now after $\phi$ integration, 
we get back to ${\mathcal{L}}_{QCD}$ since only identical 
transformations are done. However, 
{\it when the stationary phase of $\phi$ integration 
is present, it corresponds to a nontrivial phase}.
 Note that if the integral defining $Z$ is dominated by a special value of $\phi$, 
 $Z$ cannot be transformed back to ${\mathcal{L}}_{QCD}$, inequivalent 
vacuum being selected. Since $\Psi[A^a_{\mu}]$ is arbitrary, 
$\Gamma[\phi,A^a_{\mu}]$ can be regarded as the Lagrangian of the condensed phase. 
($\Psi[A^a_{\mu}]$ can be generalized to $\Psi[A^a_{\mu},\phi]$
 in (\ref{GammaA}).) We have checked our useful formula 
(\ref{GammaA}) for several model Lagrangians. In the case of QCD, 
 we know that $\phi$ condenses, so $\Gamma[\phi,A^a_{\mu}]$ has a 
 a non-trivial stationary solution corresponding to this condensate. 
Eq.(\ref{GL}) is the hydrodynamic expansion 
of $\Gamma[\phi,A^a_{\mu}]$ and we have regarded $\phi$ as a c-number field 
since $\phi$ loses fluctuations by the stationary phase mechanism. 
The stationary condition is represented by the stationary equation of
 ${\mathcal{L}}_{\phi}$ 
and we require ${\mathcal{L}}_{\epsilon,A}$ 
 not to cause the instability 
of the 
%BBB homogeneous solution satisfying
of $V'(\phi)=0$.  

\begin{center}
{\bf Normal ordering and interaction picture};~
\end{center}

Consider the expectation value of an 
arbitrary local operator $Q(A^a_{\mu}(x))\equiv Q(x)$ 
in the theory ${\mathcal{L}}_{\rm eff}$ for the homogeneous case
 of $\phi(x)=\phi$. 
%BBB the time path
%BBB running from $t_1$ to $t_2$ is represented by ${\mathcal{P}}_{t_2 ,t_1}$ and 
%BBB the state at $t=-\infty$ is written by $|\Psi\!>$. 
 Omitting the c-number ${\mathcal{L}}_{\phi}$, 
 here we introduce the $U$ operator in the path integral form by 
\begin{eqnarray*}
U(t_2 ,t_1 :\epsilon)=\int[dA^a_{\mu}]\exp\bigg(
{\rm i}\int_{t_1}^{t_2}dt d^{~\!\!3} \!x {\mathcal{L}}_{\epsilon,A}\bigg)
\end{eqnarray*}
($[dA^a_{\mu}]$ contains the factor $\sqrt{\epsilon(\phi)}$.) 
%BBB  $U$ is regarded as the operator in the following sense. 
The matrix element of $U$ between the states
 $|A^a_{\mu}(t_{1,2},\mbox{\boldmath $x$})>$ (in coordinate representation) 
 is equal to the right-hand side with $A^a_{\mu}(x)$ fixed to the value
 $A^a_{\mu}(t_{1,2},\mbox{\boldmath $x$})$ at $t_{1,2}$. 
Now we want to rewrite $U(t_2 ,t_1 :\epsilon)$ by the 
interction representation where  
``free part" is defined to be $\epsilon=1$. 
(This method of introducing the interaction representation in path-integral
 form is quite convenient since it keeps the Lorentz invariance explicitly.) 
By the homogeneity of the ststem, we take $x=0$. With $|\Psi\!\!>$ an arbitrary 
state and taking the coordinate representation in mind, let us consider
\begin{eqnarray*}
&&
\hspace{-0.3cm}<\!\!Q(0)\!\!>
=<\!\Psi|U^{\dagger}(0 ,-\infty:\epsilon)
Q(0)U(0 ,-\infty:\epsilon)|\Psi\!>
\\&&
~~~~=<\!\Psi,\phi|
U^{\dagger}(-\infty,0:1) Q(0)U(0 ,-\infty:1)
|\Psi,\phi\!>,
\\&&
|\Psi,\phi\!>\equiv
U(-\infty,0 :1)
U(0 ,-\infty:\epsilon)|\Psi\!>.
\end{eqnarray*}
In this representation, $Q$ evolves by ${\mathcal{L}}_{QCD}$ so 
 for $Q(0)=(n-4)G^{a~\!2}_{\mu\nu}(0)$,  
 the ordinary trace anomaly 
(\ref{anomaly}) holds. Here the normal order is defined with respect to the 
perturbative vacuum of $QCD$. As for the state vector, if we choose
 $|\Psi,\phi>=|0\!\!>_c$, then it is the lowest energy state in the theory 
 ${\mathcal{L}}_{QCD}$ obtaining (\ref{B}).

A comment here; if we insert the adiabatic factor connecting
 ${\mathcal{L}}_{QCD}$ and ${\mathcal{L}}_{\rm eff}$, the difference of 
the energy of the ground states of two theories (differnece of both-hand sides of 
(\ref{general})) is given by the adiabatic formula
 obtained in the interaction representation
 \cite{Gell-mann}.

\begin{center}
{\bf Color flux tube};~
\end{center}

 When color sources are present, they couple 
to the excitation field $A^a_{\mu}(x)$. 
We show here that the c-number solution of the infinite tube exists,
 where abelian components $A^3_0$ and $A^8_0$ are non-zero. This  
is sufficient for the confinement of color non-singlet state. 
Such a solution, if it exists, has a large c-number value, so it 
is stable against quantum fluctuations of other gluonic fields.
 (Quark pair creation is not discussed.) Below, the color index $a$ takes 3 or 8. 
 Now consider the case of static quark-antiquark point-like source with the strength 
$\pm g(\lambda^a /2)$ 
at $z=\pm \infty$. ($\lambda^a $ is the representation matrix of the 
quark.) 
Define $G^{a}_{k 0}\equiv \epsilon(\phi)E_k^a$ $(k=x,y,z)$, then  
div$(\epsilon(\phi)\mbox{\boldmath $E$}^a )$ is the color density.  
For an infinite tube, 
$\phi$ depends only on $\rho$ in the cylindrical coordinate $(z,\rho,\varphi)$  
and $\mbox{\boldmath $E$}^a$  
is directed along $z$ axis, which is constant over all space beause of 
rot$\mbox{\boldmath $E$}^a =0$. 
The equation of $\phi$ is  
\begin{equation}
\frac{d^2 \phi}{d^2 \rho}+
\frac{1}{\rho}\frac{d \phi}{d \rho}=
V'(\phi)-\epsilon'(\phi)E^2 /2  
\equiv |B|K'(\phi, \theta),
\label{eq}
\end{equation}
where we have written $E_z^{a} =E_z \lambda^{a} /2$, 
$E^2 =\sum_{a} E_z^{a~\!2}=E_z^2 /3$ for any color of quark. 
 We have also defined  
\[
 K(\phi,\theta)=v(\phi)- \epsilon(\phi)\theta,~~~~~~~~~\theta=E^2 /2|B|.
\]
with $v(\phi)=V(\phi)/|B|$. Note that $K(\phi,1)> -1$ by (\ref{general}).
The behavior of $K(\phi,\theta)$ is shown in the Figure.  
 \begin{figure}\hspace{5cm}
 \includegraphics{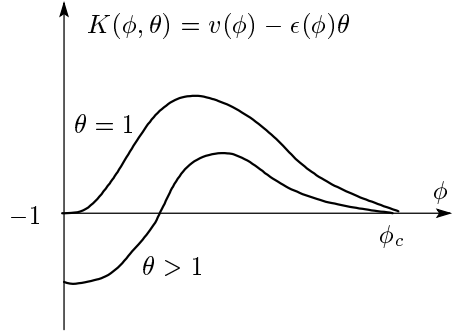}
%BBB Here is how to import EPS art
 %BBB\includegraphics{KK(phi)}% Here is how to import EPS art
 %BBB \includegraphics{K(pphi)}% Here is how to import EPS art
\caption{\label{fig:epsart}Schematic behavior of $K(\phi,\theta)$. 
 % for several values of $\theta$.  
 %AAA Note the bump structure.
 $0<\phi(0)<\phi_c$ and  
$K(\phi(0),\theta)<-1$ for $\theta>1$. Note that 
$K(0,\theta)=-\theta$.}
 \end{figure}
The solution of 
(\ref{eq}) with $\phi(\rho)$ approaching $\phi_c$ at $\rho=\infty$ is given by 
$\phi(\rho)\sim \phi_c +
 D_1 \exp(-\sqrt{W}\rho)+ D_2 \exp(\sqrt{W}\rho)$,
 with $W=|B|K''(\phi_c )$. We see $W>0$ by (\ref{general})  
 if $\alpha>1$. (When $\alpha=1$, $E<E_{\rm max}$ with $E^2_{\rm max}=
 2|B|v''(\phi_c )/\epsilon''(\phi_c )$ has to be satisfied. )
Near $\rho=0$, 
$\phi(\rho)\sim \phi(0)+C\rho^2 +F_1 \ln\rho$, with $C=|B|K'(\phi(0),\theta)/4 $. 
The requirement $F_1 =0$ and $D_2 =0$ determines $\phi(0)$ and $D_1$.   
$E$ is fixed by the total flux 
 $2\pi\int \rho d\rho \epsilon(\phi(\rho))E_z =g$.
Now we look for the solution which increases monotonically  
from $\phi(0)$ to $\phi(\infty)=\phi_c$. This requires $C>0$, which is 
shown to be consistent afterwards.   
 In order for such a solution to exist, 
$K(\phi,\theta)$ should have a bump in the region $0<\phi<\phi_c$. To see this, 
 we note that $\phi''(\rho)$ changes sign at some $\rho_d$, where $\phi''(\rho_d )=0$. 
 From (\ref{eq}), 
and $\phi'(\rho_d )>0$, $K'(\phi(\rho_d ),\theta)$ has to be positive, 
which connects smoothly to the behavior near $\phi=\phi_c$, where $K'<0$. 
So $K'=0$ at some $\phi=\phi_s$, the bump position, 
 in the range $\phi_d <\phi_s <\phi_c$. 
 If we define $\rho_s$ by $\phi(\rho_s )=\phi_s$, it is the measure of the 
radius of the tube. 
Below, $\theta>1$ is proved for any solution,
 so we need the bump structure for $\theta>1$. This condition 
is met by the existence of the bump 
for $\theta=1$, as is clear from $K(0,\theta)=-\theta$. (see the Figure. )
Thus we can say that {\it the stability of the condensed vacuum assures 
the flux tube of infinite length.} 
   
To show $\theta>1$, a sum rule is 
derived by multiplying $d\phi/d\rho$ on 
both sides of (\ref{eq}) and integrating from $\rho=0$ to $\infty$. 
Using boundary conditions on both ends, we get   
\[
\int_{0}^{\infty}\frac{1}{\rho}\bigg(\frac{d \phi(\rho)}{d\rho}\bigg)^2 d\rho
=|B|\bigg(K(\phi_c ,\theta)-K(\phi(0),\theta)\bigg)>0.
\]
Since $K(\phi_c ,\theta)=-1$, $K(\phi(0),\theta)<-1$ holds, implying $\theta>1$; 
the gain of the electric energy inside the tube is larger than the loss in breaking the 
condensation energy. If the left-hand side of (\ref{eq}) is regarded as the surface anergy, 
the difference of these two energies is sustained by the surface energy. 
If we neglect completely the surface energy, the solution satisfies $K'(\phi,\theta)=0$,
 thus $\phi=\phi_c$, or $\phi=0$. 
Setting $\phi=0$ inside the tube and $\phi=\phi_c $ outside,
 and denoting by $S$ the cross section of the tube, the energy per unit
 length $S(|B|+E^2 /2)$ is minimized 
under the constraint $SE=g$. Then $S=g/\sqrt{2|B|}$, $E=\sqrt{2|B|}$ is obtained, 
leading to $\theta=1$. The same result of course can
 be derived by the sum rule since it gives $K(\phi(0),\theta)=-1$ and  
 $\phi(0)=0$ if the surface energy is neglected. These hold 
%BBB As long as the surface energy is 
%BBB neglected, above arguments hold
 for all $\alpha\geq 1$. 
%BBBirrespective of the parameter $\alpha$.
%BBB  when we parametrize as $\epsilon(\phi)=C(\phi-\phi_c )^{2\alpha}$. 
If the surface energy is included and becomes large,
 $\theta$ and hence $E$ increases
 ($E<E_{\rm max.}$ when $\alpha=1$)
 and both the bump position and $\phi(0)$ approach $\phi_c$. 
 Thus a solution to (\ref{eq}) is assured again for any $\alpha\geq 1$. 

\begin{center}
{\bf Near the source};~
\end{center}

In the neighbourhood of the point-like colored souce of 
a quark, the condition (\ref{general}) near $\phi=0$ becomes essential. 
The equation to be solved is $\mbox{\boldmath $\nabla$}^2 \phi=
V'(\phi)-\epsilon'(\phi)E^{a~\!2}/2$.
We assume $\phi$ approaches $0$ near the source and 
search for the spherical solution with the electric field at the 
distant $r$ from the quark given by $E^a =g\lambda^a /2r^2$.
Neglecting $V'(\phi)$ and by 
 $\epsilon(\phi)\sim 1+a\phi^2$ ($a<0$)
 near $\phi=0$, 
we get 
$\phi(r)=(1/r)\exp(-\sqrt{A}/r)$, $A=-ag^2 \sum_a (\lambda^{a} /2)^2 =-ag^2 /3>0$. 
The solution is consistent with the starting assumption that 
$\phi\sim 0$ near the source. The effect of the condensation near the 
quark is smaller than any power of $r$. 

\begin{center}
{\bf Strong coupling for the canonical field};~
\end{center}

Our full quantum theory is defined by regarding (\ref{GL}) as the action which is 
functionally integrated over $\phi$ and $A^a_{\mu}(x)$. 
Let us neglect the fluctuation of $\phi$ first.  
In the condensed vacuum, since
 $\epsilon(\phi_c )=0$, $A^a_{\mu}(x)$ fluctuate indefinitely but it has no 
observable effect by the same fact $\epsilon=0$. 
 In the homogeneous case, 
%BBB physically equivalent picture is
%BBB obtained if we rewrite 
 ${\mathcal{L}}_{\epsilon,A}$ can be written 
%BBB is equivalent to QCD Lagrangian 
%BBB if it is written
 by the canonical field $\tilde{A}^a_{\mu}(x)
=\sqrt{\epsilon(\phi)}A^a_{\mu}(x)$
 (unity for the coefficient of the kinetic term).
 Then we get the same Lagrangian as QCD with 
$g$ replaced by $g_{\epsilon}=g/\sqrt{\epsilon(\phi))}$.  
Thus QCD becomes a strong coupling theory near the vacuum $\phi=\phi_c$, 
where $g_{\epsilon}\rightarrow\infty$.
Then, the dominant interaction term becomes 4-point vertex  
and by the stationary phase machanism, 
all $\tilde{A}^a_{\mu}(x)$'s
 are freezed to zero 
%BBB $\tilde{A}^a_{\mu}(x)=0$
 locally; 
gluons are confined in the operator level. 
The mechnisms of confinement of gluons
 are different for $A^a_{\mu}$ and $\tilde{A}^a_{\mu}$ but they are 
physically equivalent. 
When the quark field $q(x)$ are introduced, 
we add the usual
 term $\bar{q}(x)(\gamma^{\mu}({\rm i}\partial_{\mu}+g\mbox{\boldmath $A$}_{\mu}(x))+m)q(x)$ 
 to ${\mathcal{L}}_{\rm eff}$ with 
$\mbox{\boldmath $A$}_{\mu}
=\sum_{a=1}^8 A^a_{\mu}\lambda^a /2$. 
Note that the 
gluonic condensation is unrelated to the quark sector.   
 When $\epsilon=0$, the 
integration by $A^a_{\mu}(x)$ leads to the vanishing of the operator local 
color current of quarks; $j^a_{\mu}(x)=\bar{q}(x)\gamma_{\mu}
(\lambda^a /2)q(x)=0$ for any $a$ and $\mu$. 
 To study excitations, we note that 
 fluctuations of $\tilde{A}_{\mu}$ and
 $j_{\mu}^a$ come about only if 
$\phi$ fluctuates from the value $\phi_c$, which is governed by the glue ball 
mass. We expect $\phi\neq\phi_c$ in a 
localized region with size $1/m_g$, where $\tilde{A}^a_{\mu}$'s and 
 $j^a_{\mu}$ have non-zero values and 
 the ordinary perturbative picture works. 
Above picture of strong coupling expansion looks very similar to the same 
expansion in lattice QCD.

\begin{center}
{\bf Discussions};~
\end{center}

We try to interpret our results 
by the dual terminology. 
 The permeability
 $\mu=1/\epsilon$
 is infnite in the vacuum and for the above tube-like solution, 
the magnatic supercurrent defined by 
${\rm rot}\mbox{\boldmath $D$}^a /\epsilon=
(1/\epsilon)\mbox{\boldmath $\nabla$}\epsilon \times \mbox{\boldmath $D$}^a$
is circulating around the surface of the tube which sustains the electric flux. 
Let us introduce the dual potential 
$C_{\mu}^a$ as a Lagrange
 multiplier of the Bianchi identity \cite{Halpern,Fukuda3}
 by taking the the axial gauge 
$n^{\mu}A^a_{\mu}=0$ with some constant vector $n^{\mu}$.  
 By a simple manipulation, we get  
 the relation that ${\rm rot}\mbox{\boldmath $C$}^a$
 is equal to $\mbox{\boldmath $D$}^a$
 plus the string term which depends on $n^{\mu}$. 
Thus the dual potential $\mbox{\boldmath $C$}^a$, with the string part subtracted,  
 constitutes the tube decaying to zero for $\rho=\infty$. 
Although we have not explicitly the mass term
 $m^2 C^{a2}_{\mu}$ which breaks the magnetic gauge invariance, the  
 model discussed here behaves as dual to the 
superconductor in the above sense. 
\\
~~~Our starting formula including full quantum effcts is (\ref{GammaA}),
 which can be used in principle to determine consistently 
phenomenological parameters introduced above, 
 or to study the shielding charge due to gluons, or by including quark degrees,
the quark pair production which breaks the tube, e.t.c..

\end{document}